\documentstyle[epsbox]{elsart}

\begin{document}

\begin{frontmatter}
\title{Stochastic quantization of Born-Infeld field}
\author{Hiroshi Hotta, Mikio Namiki and Masahiko Kanenaga}
\address{Department of Physics, Waseda University, Tokyo 169, Japan}
\begin{abstract}
We\footnotemark stochastically quantize the Born-Infeld field 
which can hardly be dealtwith by means of the standard canonical 
and/or path-integral quantization methods.\footnotemark 
We set a hypothetical Langevin equation in order to quantize 
the Born-Infeld field, following the basic idea 
of stochastic quantization method. 
Numerically solving this nonlinear Langevin equation, 
we obtain a sort of ^^ ^^ particle mass'' associated with 
the gauge-invariant Born-Infeld field as a function of 
the so-called universal length. 
\end{abstract}
\end{frontmatter}
\footnotetext[1]{e-mail address:hotta@hep.phys.waseda.ac.jp}
\footnotetext[2]{
This is a revised version of which original non-electronic one 
was published in 1995 by RISE in Waseda Univ. }

\section{Introduction}
Many years ago Born and Infeld \cite{bi} presented a nonlinear 
electromagnetic field with a non-polynomial action including the 
so-called {\em universal length\/}.  One of the most important
characteristics of the Born-Infeld field is found in its static 
solution which has no divergence of static self-energy.  
Many physicists 
expected that this might be an example of divergence-free 
field theory.  However, no one could succeed to quantize the field, 
by means of the standard canonical quantization method, because of 
the complicated nonlinearity.  Even the path-integral quantization 
can hardly
be applied to this field, because we cannot easily manipulate 
such a non-polynomial action. We have to invent a new quantization 
method if we want to quantize the Born-Infeld field.

About ten years ago, Parisi and Wu \cite{pw} proposed a new 
quantization method, called {\em stochastic quantization\/}, by introducing
a hypothetical stochastic process with respect to a new (fictitious)
time, say $t$, other than ordinary time, say $x_{0}$.  
The stochastic process is so designed as to yield quantum mechanics 
in the infinite $t$-limit (thermal equilibrium limit). This theory starts
from a hypothetical Langevin equation for the stochastic process, given 
by adding the fictitious-time derivative and the random source to the classical
equation of motion.  Remember that the stochastic quantization can be
formulated only on the basis of classical field equation, without resorting to
canonical formalism.

\section{Brief review of Born-Infeld field}
The ordinary electromagnetic field is described by the following Lagrangian
density
\begin{equation}
{\cal L}=-\frac{1}{4}F_{\mu\nu}F^{\mu\nu}=\frac{1}{2}({\bf E}^{2}-{\bf H}^{2})
\ , \label{eq:eml}
\end{equation}
where $F_{\mu\nu}=\partial_{\mu}A_{\nu}-\partial_{\nu}A_{\mu}$.  
The corresponding action is given by
\begin{equation}
{\cal S}=\int d^{4}x {\cal L}\ . \label{eq:ems}
\end{equation}
Here we have followed the usual notation. Note that we keep the Minkowski
metric.

In the case of spherically symmetric static electric field, we can put
\begin{equation}
{\bf E}=-\nabla A_{0}(r), \qquad {\bf H}=0\ , \label{eq:spsym}
\end{equation}
and obtain 
\begin{equation}
A_{0}=\frac{e}{r} \label{eq:se}
\end{equation}
for a point charge $e$. That this field becomes $\infty$ for $r\rightarrow 0$ 
is a well-known fact.

Let us introduce the action functional of the Born-Infeld field \cite{bi} :
\begin{equation}
{\cal S}_{B}=\int d^{4}x \left[ -b^{2}\sqrt{1-\frac{2}{b^{2}}{\cal L}}\ 
+ b^{2} \right]\ ,  \label{eq:bis}
\end{equation}
where $1/\sqrt{b}$ is a sort of universal constant called {\em universal
length\/}, and has the dimension of length in natural unit $\hbar=c=1$.  
We can easily see that ${\cal S}_{B} \rightarrow {\cal S}$, that is,
the Born-Infeld field will become the ordinary electromagnetic field as
$b$ tends to $\infty$.  

As is well known, all physical quantities can be written only 
in terms of the dimension of length in natural unit $\hbar=c=1$.  
Many years ago Heisenberg anticipated that we could formulate 
a finite field theory, free from field-theoretical divergences, 
if we could bring a sort of {\em universal length\/} into physics 
in an appropriate way. Following his idea, Born and Infeld \cite{bi} 
proposed to use the above field given by action (\ref{eq:bis}).  
Unfortunately, however, the Heisenberg's anticipation was not 
accomplished yet even now.  

For the spherically symmetric static field $A_{0}(r)$ (\ref{eq:spsym}), the 
Born-Infeld action (\ref{eq:bis}) yields the following equation 
\begin{equation}
\frac{\partial}{\partial r} \left[ r^{2} \frac{\frac{\partial}{\partial r} 
A_{0}(r)}{\sqrt{1-\frac{1}{b^{2}}(\frac{\partial}{\partial r}A_{0}(r))^{2}}} 
 \right] =0\ , \label{eq:biseq}
\end{equation}
whose solution is given by 
\begin{equation}
A_{0}=\frac{e}{r_0}\int_{r/r_{0}}^{\infty} d\xi \frac{1}{\sqrt{1+\xi^{4}}} 
\ \rightarrow \ \left\{ 
\begin{array}{rl} 
  (e/r) & \quad \mbox{for $r \gg r_{0}$}\ , \\
  1.8541 \cdot (e/r_{0}) & \quad \mbox{for $r\rightarrow 0$}\ ,
\end{array}\right. \label{eq:bise}
\end{equation}
where $r_{0}=\sqrt{|e|/b}$.  We surely realize that the Born-Infeld field
has finite static self-energy. Of course, the self-energy goes back to
the original infinity as $b \rightarrow \infty$.

We are not interested in static field but in wave field propagating to 
remote places. The Euler-Lagrange equation of ${\cal S}_{B}$ is generally 
written as 
\begin{equation}
\frac{\delta {\cal S_{B}}}{\delta A_{\nu}(x)}=\partial_{\mu}
\left[ \frac{F^{\mu \nu}}{\sqrt{1+\frac{1}{2b^{2}}F^{2}}} \right]
=\partial_{\mu} \left[ \frac{F^{\mu \nu}}{\chi} \right]=0\ , \label{eq:bieq1}
\end{equation}
or
\begin{equation}
\partial_{\mu}F^{\mu \nu}=(\partial_{\mu} \ln \chi)F^{\mu \nu}\ , 
\label{eq:bieq2}
\end{equation}
where $F^{2}=F_{\mu \nu}F^{\mu \nu}$
 and $\chi=\sqrt{1+\frac{1}{2b^{2}}F^{2}}$.  

The right-hand side of this equation is proportional to $1/b^{2}$, and 
can be expanded in a power series of $1/b^{2}$. 
Needless to say, its unperturbed one (for $1/b=0$) is nothing 
other than the free Maxwell equation 
\begin{equation}
\partial_{\mu}F^{\mu \nu}=0\ , \label{eq:mxeq}
\end{equation}
which allows us to use the wave gauge given by 
\begin{equation}
A_{0}=0\ , \ \ \ \nabla \cdot {\bf A}=0\ .
\label{eq:wg}
\end{equation}
Consequently, one may naively expect to have the perturbative theory based on 
(\ref{eq:bieq2}) and (\ref{eq:wg}). In this case, however, we can hardly 
develop this kind of the perturbative approach to the quantized Born-Infeld
field, because the interaction part includes higher powers of derivative terms.
This is the reason why we attempt to develop 
an unperturbative approach to the quantized Born-Infeld field 
by means of {\em stochastic quantization\/} \cite{pw} in the present paper.

\section{Stochastic quantization}
As is well-known, it is convenient to use the Euclidean action ${\cal S}_{E}$ 
derived by the Wick rotation ($x_{0} \rightarrow -ix_{0}$) 
from the original Minkowski action, for the purpose of 
carrying out the Parisi-Wu stochastic quantization.  In order to perform
{\em stochastic quantization\/} of the Born-Infeld field $A_{\mu}$, 
we have to introduce
the additional dependence on fictitious-time $t$ (other than 
ordinary-time $x_{0}$) into the field quantities
and then to set the basic Langevin equation
\begin{equation}
\frac{\partial}{\partial t}A_{\mu}(x,t)=-\frac{\delta {\cal S}_{E}}
{\delta A_{\mu}}|_{A=A(x,t)}+\eta_{\mu}(x,t)
\label{eq:le}
\end{equation}
for a hypothetical stochastic process of $A_{\mu}(x,t)$ 
with respect to $t$ \cite{pw}.
Here $\eta_{\mu}(x,t)$ is the Gaussian
white-noise field subject to
\begin{eqnarray}
<\eta_{\mu}(x,t)>&=&0\ , \label{eq:st1}\\
<\eta_{\mu}(x,t)\eta_{\nu}(x',t')>&=&2\delta_{\mu \nu}
\delta(x-x')\delta(t-t') \label{eq:st2}
\end{eqnarray}
for statistical ensemble averages, where we have put $\hbar=1$ for simplicity. 

According to the prescription of stochastic quantization \cite{pw}, 
we can derive
the field-theoretical propagaters through the well-known formula
\begin{eqnarray}
\lefteqn{D^{A}_{\mu \nu}(x,x')} \nonumber\\
&& =\lim_{t\rightarrow \infty} 
\left[<A_{\mu}(x,t)A_{\nu}(x',t)> 
- <A_{\mu}(x,t)><A_{\nu}(x',t)> \right]\ , \label{eq:pro}
\end{eqnarray}
where $A_{\mu}(x,t)$ as a function of $\eta_{\mu}$ is to be obtained 
by solving (\ref{eq:le}).

Let us decompose 
$A_{\mu}$ and $\eta_{\mu}$ into their longitudinal 
and transverse components, $A^{L}_{\mu}$ and $A^{T}_{\mu}$, and 
$\eta^{L}_{\mu}$ and $\eta^{T}_{\mu}$, given by
\begin{eqnarray}
A^{L}_{\mu}&=&\frac{1}{\Box} \partial_{\mu}
\partial_{\nu}A_{\nu}\  \label{eq:al} \\
A^{T}_{\mu}&=&(\delta_{\mu \nu}-\frac{1}{\Box} \partial_{\mu}
\partial_{\nu})A_{\nu}\ ; \ \ \partial_{\mu}A^{T}_{\mu}=0\ , \label{eq:at}
\end{eqnarray}
and similar ones for $\eta$'s. Therefore, we can decompose 
the basic Langevin equation (\ref{eq:le}) as follows;
\begin{eqnarray}
\frac{\partial}{\partial t}A^{L}_{\mu}(x,t)&=&0+\eta^{L}_{\mu}(x,t)\ , \label{eq:leal} \\
\frac{\partial}{\partial t}A^{T}_{\mu}(x,t)&=&-\frac{\delta {\cal S}_{E}}
{\delta A^{T}_{\mu}}|_{A^{T}=A^{T}(x,t)}+\eta^{T}_{\mu}(x,t)\ . \label{eq:leat}
\end{eqnarray}

The absence of drift force in (\ref{eq:leal}) is an important reflection 
of the gauge invariance that ${\cal S}_{E}$ does not depend on $A^{L}_{\mu}$.
Consequently, the longitudinal component $A^{L}_{\mu}(x,t)$ makes a random
walk around its initial value $A^{L}_{\mu}(x,0)=\frac{1}
{\Box} \partial_{\mu}\phi(x)$, $\phi(x)$ 
being a scalar field. As was discussed in detail
in the case of non-Abelian gauge field \cite{nooy}, we must introduce 
{\em gauge parameter\/} $\alpha$ by taking average of $\phi$ over 
random fluctuations around zero ($\overline{\phi(x)\phi(x')}
=\alpha \delta(x-x'))$. Thus we obtain
\begin{equation}
D^{A^L}_{\mu \nu}(x,x')=\frac{1}{\Box} \partial_{\mu}
\partial_{\nu}(\alpha \frac{1}{\Box}+2t)\delta(x-x')\ . \label{eq:alpr}
\end{equation}
Of course, we know that the longitudinal components never appear
in gauge-invariant (physical) quantities.

We should also notice that the transverse components and their propagaters
are completely decoupled with the longitudinal ones in the present case. 
This is an important point quite different from the non-Abelian gauge field
case. Thus we can safely discard $A^{L}_{\mu}$, and use
the wave gauge (\ref{eq:wg}), even in the present case, for the purpose 
of deriving propagaters of the transverse components. 

Considering wave propagation along the $z=x_{3}$-axis, we put
\begin{eqnarray}
A_{0}&=&0\ , \qquad A_{3}=0\ , \label{eq:a03} \\
A_{1}&=&A_{1}(x_{0},x_{3})\ , \qquad
A_{2}=A_{2}(x_{0},x_{3})\ , \label{eq:a12} 
\end{eqnarray}
and
\begin{eqnarray}
\eta_{0}&=&0\ , \qquad \eta_{3}=0\ , \label{eq:eta03} \\
\eta_{1}&=&\eta_{1}(x_{0},x_{3})\ , \qquad
\eta_{2}=\eta_{2}(x_{0},x_{3})\ . \label{eq:eta12} 
\end{eqnarray}
Note that $A_{\mu}$ and $\eta_{\mu}$ have no longitudinal components.

In this case, the Euclid action becomes 
\begin{equation}
{\cal S}_{E}=\int dx_{0}dx_{3} \left[ b^{2}\sqrt{1+\frac{1}{b^{2}}{F_{E}^{2}}}
-b^{2} \right]\ ,
\label{eq:es2}
\end{equation}
where $F_{E}^{2}=(\partial_{0} A_{1})^{2}+(\partial_{3} A_{2})^{2}
+(\partial_{0} A_{2})^{2} +(\partial_{3} A_{1})^{2}$. 
The corresponding classical field equation is given by
\begin{equation}
\frac{\delta S_{E}}{\delta A_{i}(x)}=-\partial_{0} \left[ 
\frac{\partial_{0} A_{i}}{\sqrt{1+\frac{1}{b^{2}}F_{E}^{2}}} \right]
-\partial_{3} \left[ 
\frac{\partial_{3} A_{i}}{\sqrt{1+\frac{1}{b^{2}}F_{E}^{2}}} \right]
\ =0, \ \ (i=1,2)\ .
\label{eq:cleqwg}
\end{equation}
Note that the dimensions of the field and $b$ are different from 
the original ones: $[A_{i}]=[L^{0}]$ and $[b^{-1}]=[L]$, $[L]$ standing for
the dimension of length. 

Based on the classical equation, we can set up the basic Langevin equation
for stochastic quantization of the Born-Infeld field as follows;
\begin{eqnarray}
\frac{\partial}{\partial t}A_{i}(x_{0},x_{3},t)&=&
\partial_{0} \left[ 
\frac{\partial_{0} A_{i}}{\sqrt{1+\frac{1}{b^{2}}F_{E}^{2}}} \right]
+\partial_{3} \left[ 
\frac{\partial_{3} A_{i}}{\sqrt{1+\frac{1}{b^{2}}F_{E}^{2}}} \right]
+\eta_{i}(x_{0},x_{3},t)\ , \nonumber \\
(i=1,2)\ , \label{eq:lgeqwg}
\end{eqnarray}
where the fluctuating source-field $\eta_{i}$ should have the
statistical properties
\begin{eqnarray}
&& <\eta_{i}(x_{0},x_{3},t)>_{\eta} = 0\ , \label{eq:tks1} \\
&& <\eta_{i}(x_{0},x_{3},t)\eta_{j}(x_{0}',x_{3}',t')>_{\eta} = 2\delta_{ij}
\delta(x_{0}-x_{0}')\delta(x_{3}-x_{3}')\delta(t-t')\ . \label{eq:tks2}
\end{eqnarray}
Consequently, we have to obtain 
$A_{i}$ as a function (or functional) of
$\eta_{i}$, by solving (\ref{eq:lgeqwg}), and then to calculate expectation 
values of physical quantities, by making use of (\ref{eq:tks1}) and
(\ref{eq:tks2}). For example, the field-theoretical propagater of $A_{i}$
is given by the formula
\begin{eqnarray}
\Delta_{ij}^{A}(x_{0}-x_{0}',x_{3}-x_{3}') && \equiv 
\lim_{t \rightarrow \infty} 
\left[ <A_{i}(x_{0},x_{3},t)A_{j}(x_{0}',x_{3}',t)> \right. \nonumber\\
&& \mbox{   } \left. -<A_{i}(x_{0},x_{3},t)><A_{j}(x_{0}',x_{3}',t)> \right].
\label{eq:prgt} 
\end{eqnarray}

For conventional fields, we can extract real information about the particle
mass, ${\cal M}$ and ${\cal M}'$, associated with the field 
(or the first energy gap) from the asymptotic formulas
\begin{eqnarray}
\Delta_{ii}^{A}(0,x_{3})& \stackrel{|x_{3}| \rightarrow \infty}{\longrightarrow} &\mbox{const.} \exp[-{\cal M}|x_{3}|] \ , \label{eq:m1} \\
\Delta_{ii}^{A}(x_{0},0)& \stackrel{|x_{0}| \rightarrow \infty}{\longrightarrow} &\mbox{const.} \exp[-{\cal M}'|x_{0}|] \label{eq:m2}
\end{eqnarray}
($i$: no summation), in which we can put ${\cal M}={\cal M}'$ 
for the Euclidean symmetry
in space-time. Unfortunately in the Born-Infeld case, however, 
we have no reliable theory to justify the procedure (\ref{eq:m1}) and/or (\ref{eq:m2}) to give mass. Despite of this situation, 
we intend to follow the conventional approach to the ^^ ^^ particle mass'' 
associated with the (transverse) Born-Infeld field, 
based on (\ref{eq:m1}) 
and/or (\ref{eq:m2}).

\section{Numerical simulation and particle mass}
Needless to say, we know that it is very difficult to solve (\ref{eq:lgeqwg})
analytically, so that we are inevitably enforced to deal with it by means of 
numerical simulation.  For this purpose, we 
first discretize the Langevin equation (\ref{eq:lgeqwg}) 
on an $N\times N$ lattice with spacings $\Delta x_{0}$ and $\Delta x_{3}$ 
(along time and space directions, respectively).
Denoting the Born-Infeld field on the $(k,l\/)$-th lattice point 
by $A_{i;k,l}(t)$, where $i$ stands for the $i$-th component 
of the field and $k,l$ for the ordinary time and spatial 
position, then we write down the discretized Langevin 
equation as 
\begin{eqnarray}
\lefteqn{\frac{A_{i;k,l}(t + \Delta t) - A_{i;k,l}(t)}{\Delta t}} 
\nonumber\\
&& = \frac{G_{i;k+1,l}(t) - G_{i;k,l}(t) }{\Delta x_0}
    +\frac{H_{i;k,l+1}(t) - H_{i;k,l}(t)}{\Delta x_3}
    +\sqrt{\frac{2}{\Delta x_{0} {\Delta x_{3}} {\Delta t}}}
     N_{i;k,l}(t) \ , \nonumber\\
&& \label{eq:disl1}
\end{eqnarray}
where 
\begin{eqnarray}
G_{i;k,l}(t)
&=& \frac{ \frac{1}{\Delta x_{0}} ( A_{i;k,l}(t) - A_{i;k-1,l}(t) )}
        {\sqrt{ 1 + F_{E}^{2} }} \ ,\label{eq:g} \\
H_{i;k,l}(t)
&=& \frac{ \frac{1}{\Delta x_{3}} ( A_{i;k,l}(t) - A_{i;k,l-1}(t) )}
        {\sqrt{ 1 + F_{E}^{2} }} \label{eq:h}
\end{eqnarray}
with 
\begin{eqnarray}
F_{E}^{2}
&=&\frac{1}{b^{2}} \left\{ \left[ \frac{A_{1;k,l}(t) - A_{1;k-1,l}(t)}{\Delta x_0} \right]^{2} 
 +\left[ \frac{A_{2;k,l}(t) - A_{2;k-1,l}(t)}{\Delta x_0} \right]^{2} \right.
\nonumber \\
& & \left. +\left[ \frac{A_{1;k,l}(t) - A_{1;k,l-1}(t)}{\Delta x_3} \right]^{2} 
 +\left[ \frac{A_{2;k,l}(t) - A_{2;k,l-1}(t)}{\Delta x_3} \right]^{2} \right\} \label{eq:disl3}
\end{eqnarray}
for drift terms, and
\begin{eqnarray}
& &
 < N_{i;k,l}(t) >_{N} = 0,\qquad
 < N_{i;k,l}(t) N_{i;k',l'}(t') >_{N}
  = \delta_{ij} \delta_{kk'} \delta_{ll'} \delta_{tt'}\ ,\label{eq:disn1}\\
& &(i, j = 1, 2), \qquad (l, k = 1, \ldots, N)  \nonumber
\end{eqnarray}
for noise terms. 

Here let us introduce a scale unit $a$, which has dimension of length, 
and put relevant quantities in the following way:
\begin{eqnarray}
& &{\Delta x_{0}}={\Delta {\tilde x}_{0}}a,\ 
   {\Delta x_{3}}={\Delta {\tilde x}_{3}}a,\ 
   {\Delta t}={\Delta {\tilde t}}{a^{2}},\ 
   b={\tilde b} a^{-1}, \ \nonumber \\
& &A_{i;k,l}(t)={\tilde A}_{i;k,l}(\tilde t),\ 
G_{i;k,l}(t)={\tilde G}_{i;k,l}(\tilde t){a^{-1}},\ \nonumber \\ 
& &H_{i;k,l}(t)={\tilde H}_{i;k,l}(\tilde t){a^{-1}},\ 
N_{i;k,l}(t)={\tilde N}_{i;k,l}(\tilde t),\ \nonumber \\ 
& &{\cal M}=\tilde{{\cal M}} a^{-1}, \label{eq:mass}
\end{eqnarray}
Note that all quantities with tilde are dimensionless.
Thus, the equations (\ref{eq:disl1}) and (\ref{eq:disn1}) are
rewritten as 
\begin{eqnarray}
& &
\frac{{\tilde A}_{i;k,l}({\tilde t} + \Delta {\tilde t}) 
- {\tilde A}_{i;k,l}(\tilde t)}{\Delta {\tilde t}} \nonumber\\
& &=\frac{{\tilde G}_{i;k+1,l}(\tilde t) - {\tilde G}_{i;k,l}(\tilde t)}
{ \Delta {\tilde x}_0 } 
+\frac{{\tilde H}_{i;k,l+1}(\tilde t) - {\tilde H}_{i;k,l}(\tilde t) }
{ \Delta {\tilde x}_3 }
+\sqrt{
\frac{2}{ {\Delta {\tilde x}_{0}}{\Delta {\tilde x}_{3}}{\Delta {\tilde t}} }
} {\tilde N}_{i;k,l}(\tilde t)\ ,\nonumber \\
&& \\
& &< {\tilde N}_{i;k,l}(\tilde t) >_{\tilde N} = 0, \qquad  
< {\tilde N}_{i;k,l}(\tilde t) 
{\tilde N}_{i;k',l'}({\tilde t}') >_{\tilde N} 
= \delta_{ij} \delta_{kk'} \delta_{ll'} \delta_{{\tilde t}{\tilde t}'}\ .
\end{eqnarray}
$\tilde G_{i;k,l}$ and $\tilde H_{i;k,l}$ include the dimensionless 
$\tilde F_{E}^{2}$ in the same way as $G_{i;k,l}$ and $H_{i;k,l}$
depend on $F_{E}^{2}$.  
Note that only $\tilde F_{E}^{2}$ contains 
the Born-Infeld parameter $\tilde b$.

In the conventional field theory, this scale unit $a$ can be determined
by making use of the renormalization group theory. In the present case, 
however, we have no reliable theory to determine 
the scale unit, and then we shall inevitably calculate all quantities 
(in particular, the ^^ ^^ particle mass'') on an arbitrary scale. 
Only for the sake of simplicity, let us put $a=1$, in order to go on our 
procedure. 
For a while from now on, 
we suppress those tilders which are put on the quantities.
Therefore, the ^^ ^^ particle mass'', ${\cal M}$, will be given as 
a dimensionless quantity in this scheme.

In order to solve numerically the above equation and obtain 
the field-theoretical propagater $\Delta_{ij}(x_{0},x_{3})$ on the above lattice, 
we should introduce the periodic boundary condition given by
\begin{equation}
A_{i;k,l+2l_{c}} = A_{i;k,l}\ ,\  A_{i;k+2k_{c},l}=A_{i;k,l} \ ,
\label{eq:bdc}
\end{equation}
where $l_{c}, k_{c}$ and $2l_{c}, 2k_{c}$ stand for the lattice center 
and the period, respectively.  
Practically, we have used the Langevin-source method
(for example, see \cite{pw}), in which we have performed $5.4 \times 10^6$ 
iterations for a lattice of $20 \times 20$ sites with
$\Delta t= 0.01$ and $l_{c}=k_{c}=10$ (to realize thermal equilibrium), 
and then use the subsequent $2.0 \times 10^5$ 
iterations to calculate the field-theoretical propagater.  

\begin{figure}
\epsfile{file=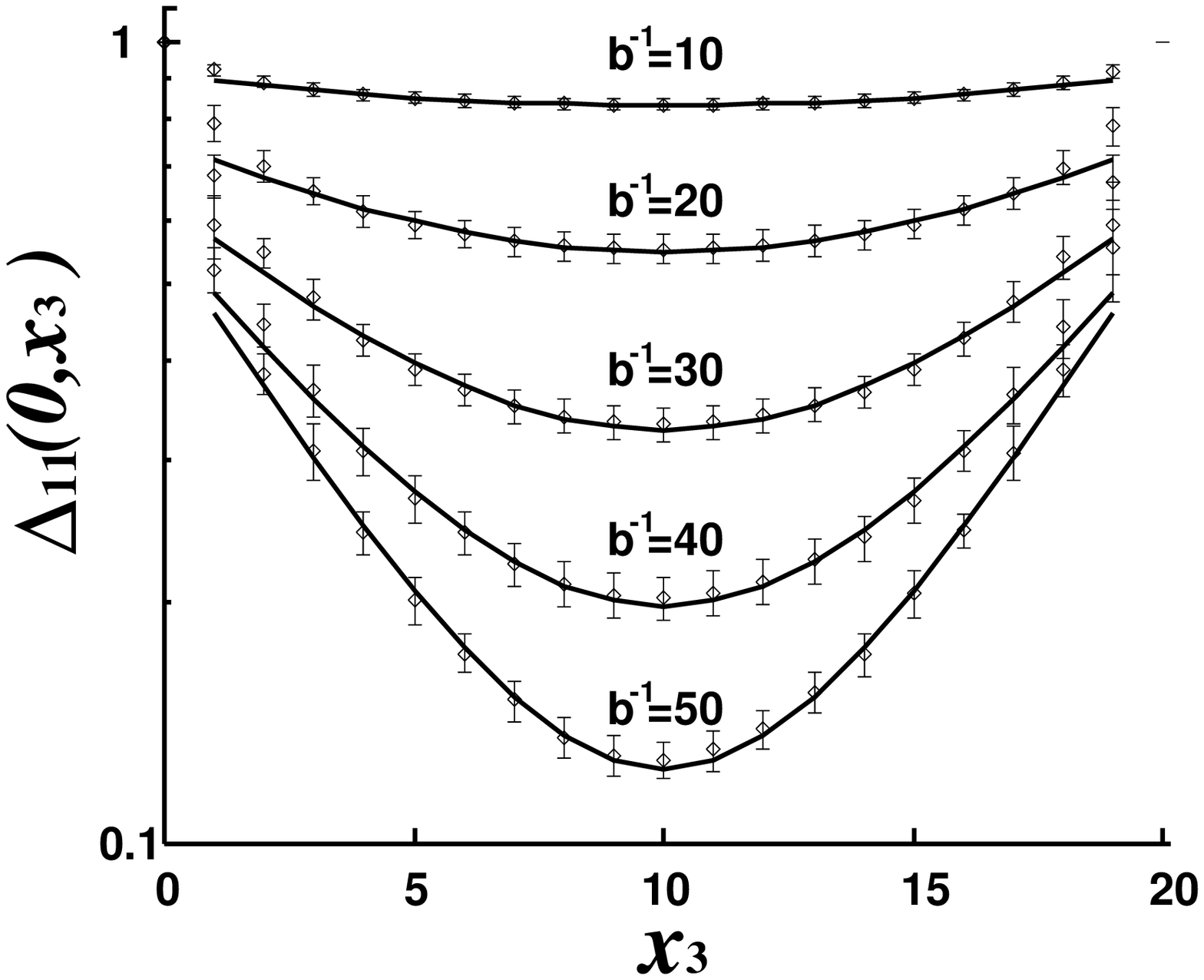,hscale=0.60,vscale=0.5}
\caption{Field-theoretical propagaters for several values of
$b^{-1}$.}
\end{figure}
Figure 1 shows our numerical results of the field-theoretical 
propagaters 
$\Delta_{11}(0,x_{3})$ for 
$b^{-1}=10,20,30,40$ and $50$, together with curves given by 
the asymptotic formula
\begin{equation}
{\Delta}_{ii}(X,0)={\Delta}_{ii}(0,X)=C\frac{\cosh {\cal M}|X-x_{c}|}
{\cosh {\cal M}|x_{c}|}\ \ (i: \mbox{no summation}),\ \label{eq:asm}
\end{equation}
where $x_{c}$ stands for the center of lattice.  
Also we numerically have got similar field-theoretical propagaters 
for other directions and/or components.  
Equation (\ref{eq:asm}) is the substitute of (\ref{eq:m1}) 
and/or (\ref{eq:m2}) 
under the boundary condition (\ref{eq:bdc}).  We have estimated the 
^^ ^^ particle mass'', ${\cal M}$, associated with the Born-Infeld field, 
by making use of $\chi^{2}$-fitting based on (\ref{eq:asm}).
As shown in Table 1, our results are the 
following: ${\cal M} = 0.0415, 0.0835, 0.1260, 0.1693, 0.2162$, correspondingly 
to $b^{-1} = 10, 20, 30, 40, 50$, where $\chi^2$ is $4.69 \times 10^{-4}$.
We have estimated the statistical fluctuations for the fictitious time
as accuracies in Table 1. Here we have
put $C={\Delta}_{ii}(0,0)$ $=<A_{i}^{2}>$ ($i$: no summation). Note that $C$ 
is independent of $i$ due to the space-time uniformity, and that $C$ is 
gauge-invariant.

Rigorously speaking from the point of view of (\ref{eq:mass}),
we can only assert that
the above ${\cal M}$ is proportional to the ^^ ^^ particle mass''
associated with the (transverse) Born-Infeld field.  We should repeat that 
we have no renormalization group theory to give the scaling formula 
in the case of Born-Infeld field. Remember that the problem is 
still open to questions. 
In this paper, however, we are talking about the ^^ ^^ particle mass'' 
by ${\cal M}$ which is given by (\ref{eq:asm}).

\begin{table}
\caption{The ^^ ^^ particle mass'', ${\cal M}$, associated with the 
Born-Infeld field}
\begin{tabular}{c|ccccc} \hline
$b^{-1}$ & 10 & 20 & 30 & 40 & 50 \\ \hline
${\cal M}$ & 0.0415 & 0.0835 & 
0.1260 & 0.1693 & 0.2162 \\ 
Accuracy & $\pm 0.0062$ & $\pm 0.0021$ & 
$\pm 0.0017$ & $\pm 0.0017$ & $\pm 0.0015$ \\ \hline
\end{tabular}
\end{table}

\begin{figure}
\vspace{1cm}
\epsfile{file=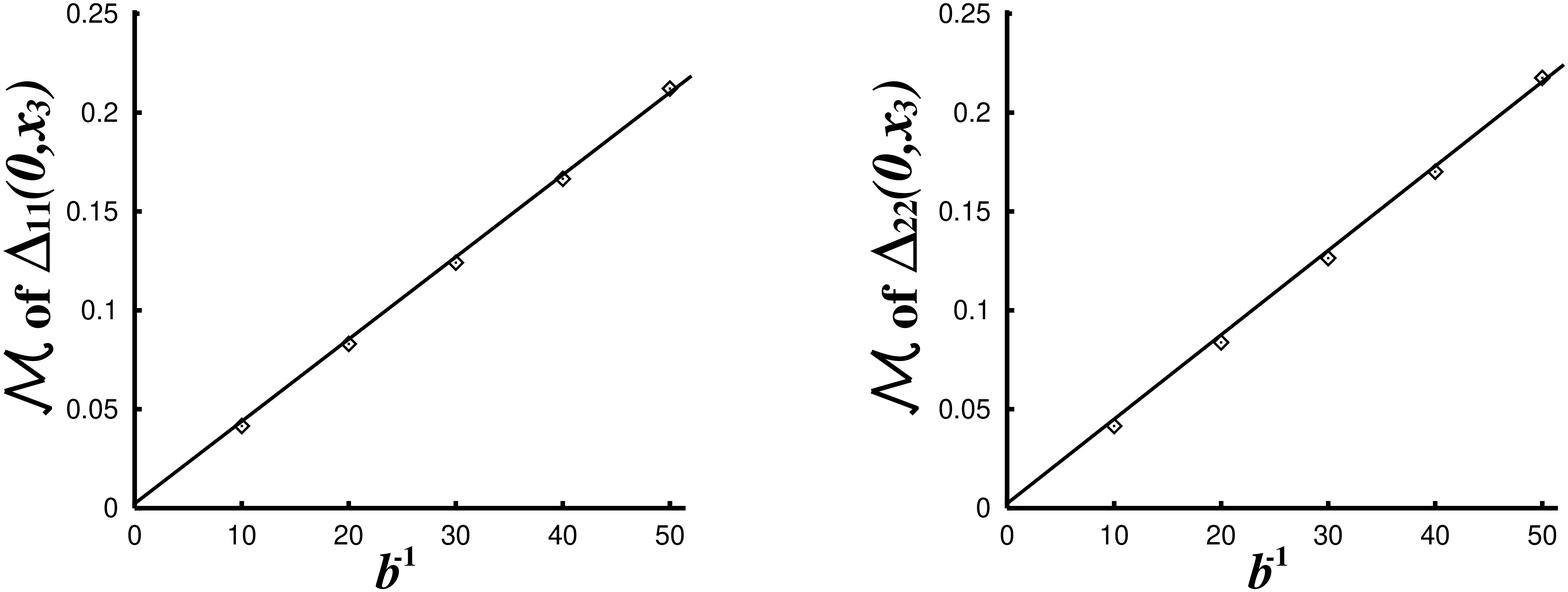,hscale=0.3,vscale=0.3}
\caption{^^ ^^ Particle mass'', ${\cal M}$, as a function of $b^{-1}$.}
\end{figure}

\begin{figure}
\epsfile{file=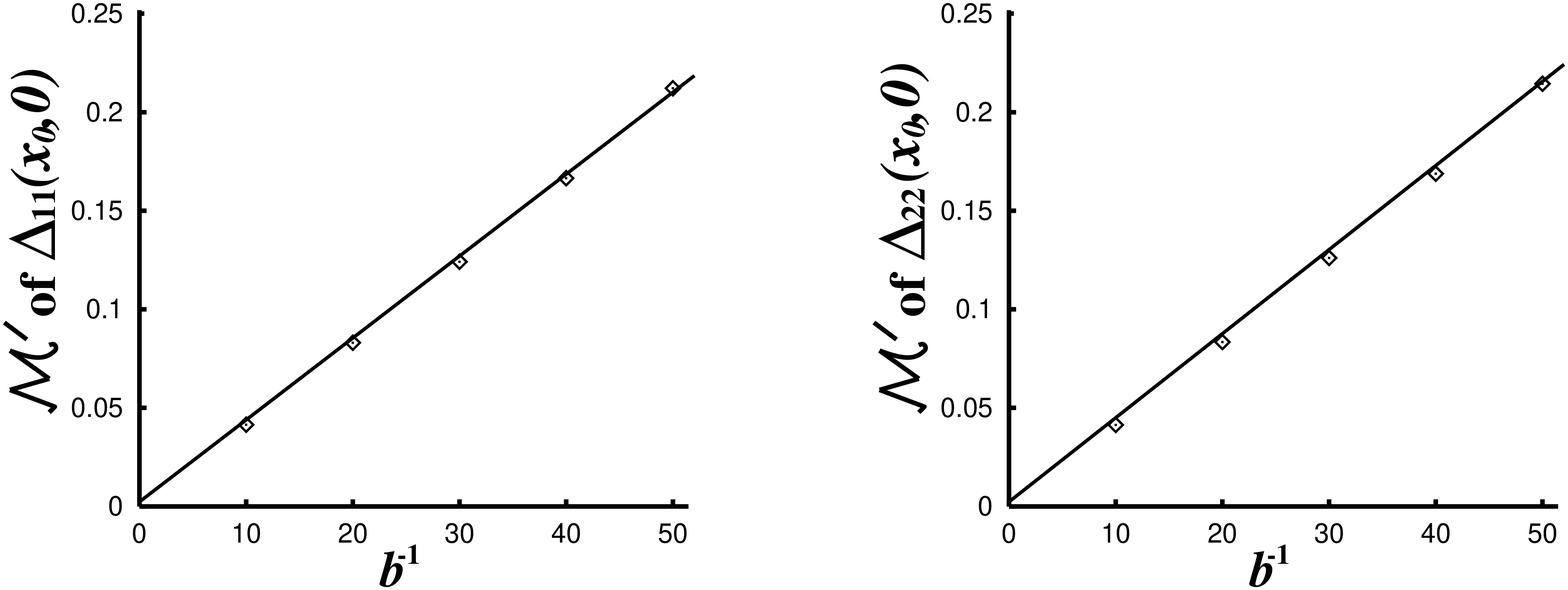,hscale=0.3,vscale=0.3}
\caption{^^ ^^ Particle mass'', ${\cal M'}$, as a function of $b^{-1}$.}
\end{figure}
Figures 2 and 3 plot the ^^ ^^ particle mass'', given by ${\cal M}$ 
and ${\cal M'}$, as a function of $b^{-1}$.  Observe that ${\cal M}={\cal M'}$.
It seems that the ^^ ^^ particle mass'' is proportional to $b^{-1}$, 
but unfortunately, we do not know what kind of physical 
implications this fact suggests.  Another important point should be that 
the ^^ ^^ particle mass'' seems vanishing, in the case of $b^{-1}=0$, 
as expected from the fact that the Born-Infeld field must go back 
to the free Maxwell field in this limit.  All results are presented 
in Table 1.  

Table 1 also tells us that the (dimensionless) ^^ ^^ particle mass'' 
on this scale (with $a=1$) distributes
over a very small region.  

Table 1 or Figures 2 and 3 can be fitted well 
by a single formula given by 
\begin{equation}
{\cal M} = \gamma \frac{1}{b}\ , \qquad \gamma = 0.00426 \ .
\end{equation}
This equation is rewritten in terms of 
${\cal M}$ and $b^{-1}$ having dimension as 
\begin{equation}
{\cal M}
 = (\frac{\gamma}{a^2}) \frac{1}{b}\ , \qquad \gamma = 0.00426 \ .
\end{equation}

Here let us try to choose $a=b^{-1}$ as the scale unit, 
then we obtain 
\begin{equation}
{\cal M} = \gamma b\ , \qquad \gamma = 0.00426 \ ,
\end{equation}
or 
\begin{equation}
\tilde{{\cal M}} = \gamma , \qquad \gamma = 0.00426 \ ,
\end{equation}
in the dimensionless expression. 
because $\tilde{b} = 1$ in this case.  
This implies that the constant $\gamma$ is nothing 
other than the ^^ ^^ particle mass'',
being independent of the {\em universal length\/}, 
on the scale adjusted by $a=b^{-1}$.  

Finally, let us examine whether our ^^ ^^ particle mass'' 
${\cal M}$ can be regarded as a sort of particle mass 
in the sense of conventional field theory.  
For this purpose, we should numerically compute 
Fourier transform ${\tilde \Delta}_{ii} (k^{2})$ 
given by 
\begin{equation}
{\tilde \Delta}_{ii}(k^{2})
 =\int_{0}^{N} \frac{dx_{0}}{\sqrt{N}}
  \int_{0}^{N} \frac{dx_{3}}{\sqrt{N}}
   e^{-ik_{0}x_{0}-ik_{3}x_{3}} 
    \Delta_{ii}(x_{3},x_{0})\ , \label{eq:ft}
\end{equation}
$N$ standing for lattice size.  
If ${\cal M}$ meant a sort of particle mass 
in this sense, we could hardly observe so sharp 
${\cal M}$-dependence of ${\tilde \Delta}_{ii} (k^{2})$ 
as a function of $k^{2}$, for 
$\sqrt{k^2} \gg {\cal M}=0.0415, 0.0835, 0.1260, 0.1693, 0.2162$.  
\begin{figure}
\epsfile{file=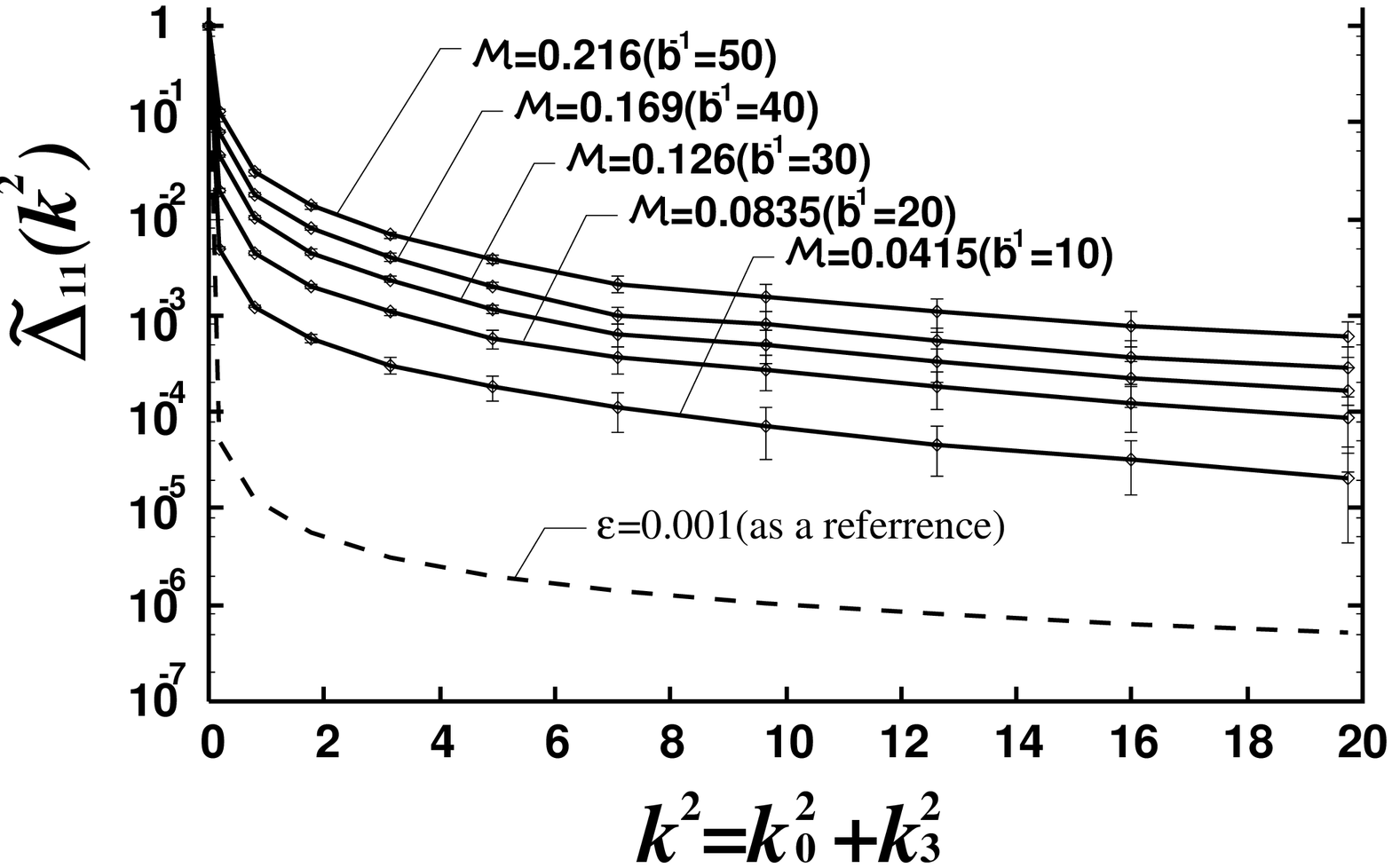,hscale=0.6,vscale=0.55}
\caption{${\tilde \Delta}_{11}(k^{2})$ corresponding to masses obtained in Table 1}
\end{figure}
Figure 4 shows our numerical results (see \cite{ni} for 
technical details), in which all curves for various 
$b$'s are normalized to ${\tilde \Delta}_{ii}(0)=1$.  
We can clearly observe in Fig.~4 that ${\tilde \Delta}_{ii}(k^{2})$ 
is almost independent of $k^{2}$, except in its height.  
That is to say, our anticipation seems justified.  
For comparison, 
we put the dashed curve representing a free-propagater (being 
a substitute of massless free Maxwell field), $\epsilon^{2}/
(k^{2} + \epsilon^{2})$, with a very small mass 
$\epsilon=0.001 \ll {\cal M}$.

In order to reconfirm this fact, we present Fig.~5,
(for $\frac{1}{{\cal M}^2}\Delta_{ii}(k^2)$ versus $k^2$) 
stressing that all curves overlap each others for 
larger $k^{2}$.  
Note that long distance behavior of the propagater 
in configuration space gives ${\cal M}$, 
while short distance behavior determines Fourier transforms
for larger $k^2$.
\begin{figure}
\epsfile{file=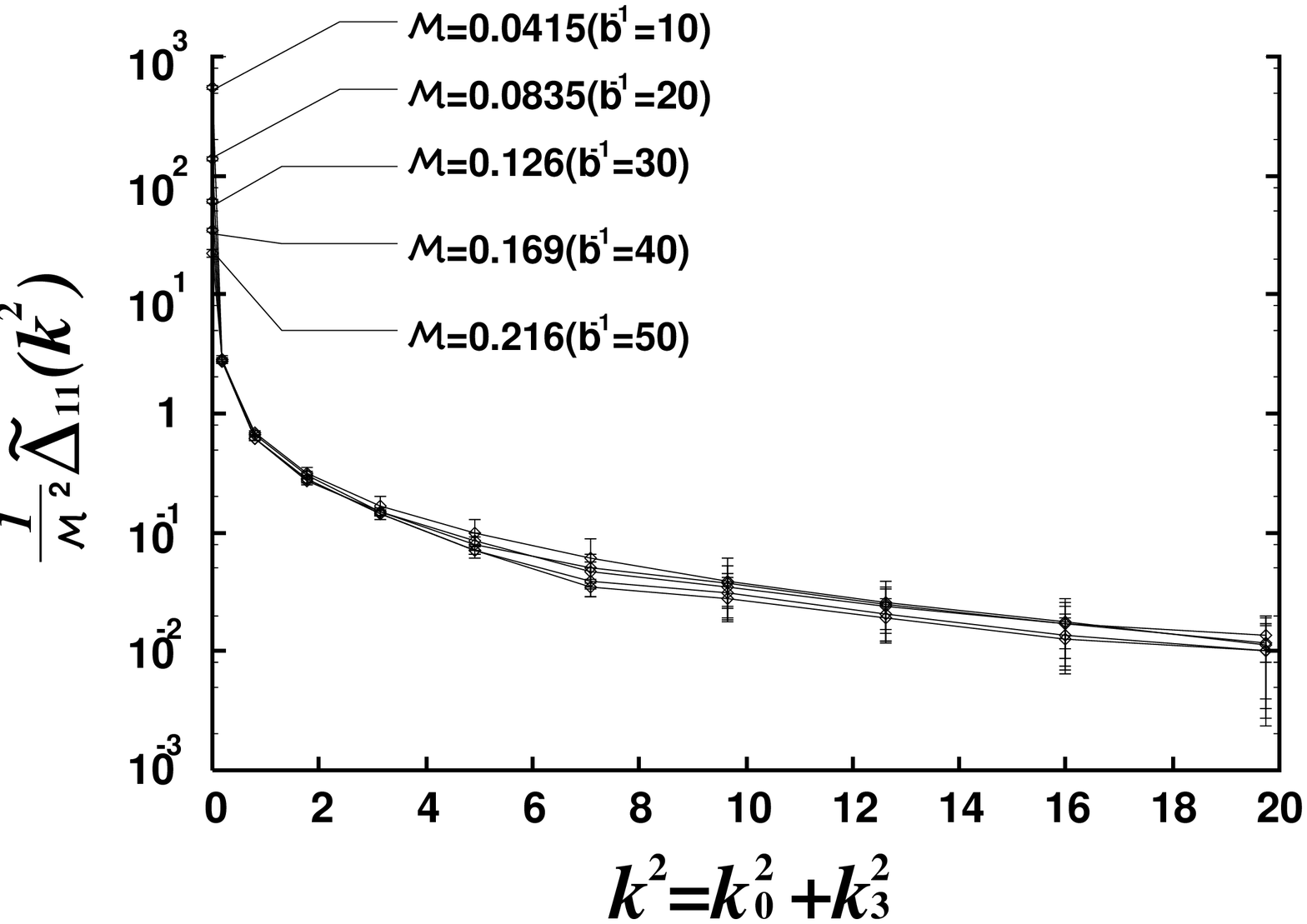,hscale=0.6,vscale=0.55}
\caption{$\frac{1}{{\cal M}^2} {\tilde \Delta}_{ii}(k^{2})$ versus $k^{2}$}
\end{figure}

Moreover, we compare a special ${\tilde \Delta}_{ii}(k^{2})$ 
with the corresponding
Feynman propagater ${\cal M}^{2}/(k^{2}+{\cal M}^{2})$ 
for ${\cal M}=0.126$ in Fig.~6, as an example.  
In this figure one could observe that the difference 
between them would represent a possible (damping) effect due to 
the non-linearity of the Born-Infeld field.  

\begin{figure}
\epsfile{file=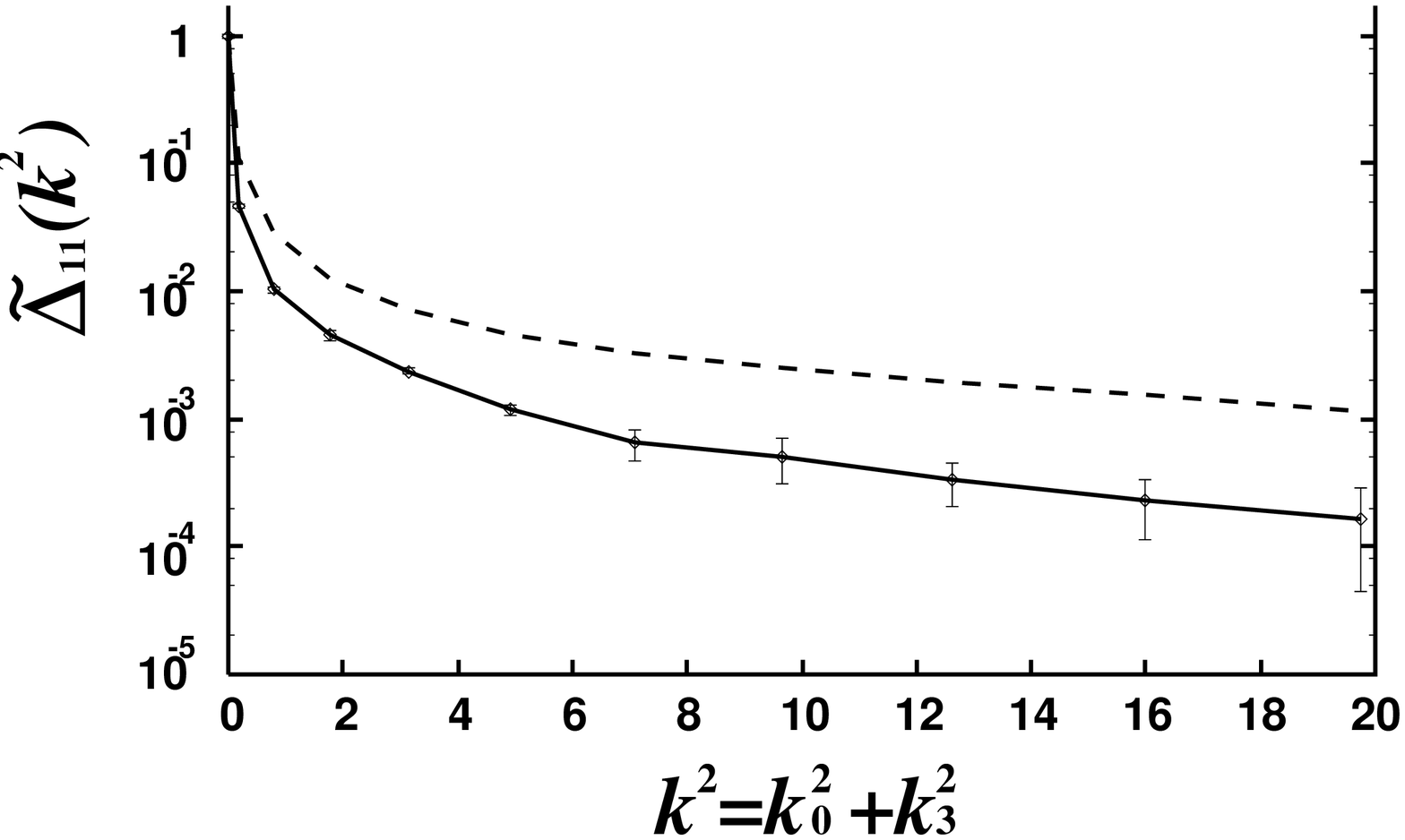,hscale=0.6,vscale=0.55}
\caption{Comparison of ${\tilde \Delta}_{11}(k^{2})$ with ${\cal M}^{2}/(k^{2}+
{\cal M}^{2})$ for ${\cal M}=0.126$.}
\end{figure}

\section{Conclusion}
Summarizing, we have stochastically quantized the Born-Infeld field,
characterized by the so-called {\em universal length\/},
which cannot be dealt with by means of the conventional quantization methods.
Even though we can hardly justify the whole procedure theoretically,
we have derived the ^^ ^^ particle mass'' associated with the (transverse) 
Born-Infeld field, as a function of the {\em universal length\/},
through the conventional formulas to give them.

It would be interesting to observe that we have derived 
the ^^ ^^ particle mass'' from a perfectly gauge-invariant
field theory. Of course, we can guess that the ^^ ^^ particle mass'' 
is produced by introducing the {\em universal length\/} $b^{-1}$ 
having the dimension of length.

The authors are indebted to Drs. I. Ohba,
 S. Tanaka, Y. Yamanaka, K. Okano and B. Zheng 
for many discussions and suggestions.

\end{document}